\documentstyle[prl,aps]{revtex}
\draft
\input{psfig.sty}
\begin{document}
\twocolumn[\hsize\textwidth\columnwidth\hsize\csname@twocolumnfalse\endcsname
\title{Nucleation of vortices by rapid thermal quench}
\author{I.S. Aranson$^{1}$, N.B. Kopnin$^{1,2}$ and  V.M. Vinokur$^{1}$}
\address{$^{1}$Argonne National Laboratory, 9700 
South Cass Avenue, Argonne, IL 60439}
\address{$^2$ L.D. Landau Institute for Theoretical Physics, 
117334 Moscow, Russia} 
\date{\today}

\maketitle
\begin{abstract}
We show that vortex nucleation in 
superfluid $^3$He by rapid thermal quench in the presence of superflow
is dominated by a transverse instability of the moving 
normal-superfluid interface. 
Exact expressions for the instability threshold 
as a function of supercurrent density and the front velocity 
are found. The results are verified by numerical solution
of the  Ginzburg-Landau equation. 
\end{abstract}


\narrowtext
\vskip1pc]

Formation of topological defects under 
a rapid quench is a fundamental problem of
contemporary physics
promising to shed a new light on the early stages of the evolution of 
the Universe. For  homogeneous cooling a fluctuation-dominated 
formation mechanism has been suggested by Kibble and Zurek 
(KZ)\cite{Kibble,Zurek,zurek}.  Normally,  cooling is associated with  
an inhomogeneous temperature distribution
accompanied by  a phase separating interface which moves 
through the system as temperature decreases.  A generalization 
of the KZ scenario was suggested in Ref. \cite{kibbleVol} for 
inhomogeneous phase transitions in superfluids: if the front
moves faster than the normal--superfluid interface
a large supercooled region which is left behind becomes unstable
towards fluctuation-induced nuclei.

Superfluid $^3$He offers a unique ``testing ground''  for 
rapid phase transitions \cite{ekv}. 
Recent experiments where a rotating superfluid $^3$He was locally heated
well above the 
critical temperature by absorption of neutrons \cite{explos} 
revealed vortex formation under a rapid second--order phase transition.
The TDGL analysis was applied to study a
propagating normal--superfluid interface under inhomogeneous cooling
\cite{kopnin} and the formation of a large supercooled region was confirmed. 
The fluctuation--dominated  
mechanism may thus be responsible  for creation of initial vortex loops. 
It is commonly accepted 
that these  initial vortex loops are further inflated by the
superflow and give rise to a macroscopic number of large vortex lines 
filling the
bulk superfluid.

In this Letter we report a novel mechanism of vortex formation 
which overtakes growth of the initial  loops appearing in the supercooled
region. We study the  entire  process of vortex formation   
in the presence of 
a superflow using  TDGL dynamics.  We take into account the temperature
evolution 
due to thermal diffusion. We find  analytically and 
confirm by numerical simulations that the normal--superfluid
interface becomes unstable with respect to transverse undulations 
in the presence of a superflow. 
These undulations  quickly transform into large primary 
vortex loops which then separate themselves from the interface. 
Simultaneously,
a very large number of small secondary vortex/antivortex nuclei are created 
in the supercooled
region by fluctuations resembling the conventional KZ mechanism.
The primary vortex loops 
screen out the superflow in the inner region causing the annihilation of 
the secondary  vortex/antivortex  nuclei.   
The number of {\it surviving} vortex loops 
is thus much smaller than what
anticipated from the KZ conjecture.

{\it Model}. -- 
We use the TDGL  model for a scalar order parameter $\psi$
ignoring the non--relevant  complexity of the $^3$He--specific
multicomponent
order parameter:
\begin{equation} 
\partial_t \psi = \Delta \psi + (1 - f({\bf r },t) ) \psi -|\psi| ^2 \psi
+\zeta({\bf r},t).
\label{gle1}
\end{equation} 
Here 
$\Delta $ is the three-dimensional (3D) Laplace operator. 
Distances and time are measured 
in units of the coherence length $\xi (T_\infty )$ and the 
characteristic  time $\tau_{GL} (T_\infty )$, respectively.
These quantities are taken at temperature $T_\infty$ far from the heated
bubble.
The local temperature is controlled by heat diffusion and
evolves as  
$f({\bf r },t) =  E_0  \exp(-r^2 /\sigma t)t^{-3/2}$ 
where $\sigma$ is the normalized diffusion coefficient.
$E_0 $ determines the initial temperature of the hot 
bubble $T^*$ and is 
proportional to the deposited energy ${\cal E}_0$ such that
$E_0={\cal E}_0/\left[ C(T_c-T_\infty) 
\xi ^3(T_\infty )(\pi \sigma )^{3/2}\right]$ where
$C$ is the heat capacity.  Since the deposited energy
is large compared to 
the
characteristic superfluid energy, we assume  $E_0 \gg 1$. 
The time at which the temperature in 
the center of the
hot bubble drops down to $T_c$ is
$t_{\rm max}=E_0^{2/3}$.
The Langevin force
$\zeta$  with the correlator 
$\langle \zeta \zeta^\prime \rangle=
2 T_f \delta({\bf r -r^\prime}) \delta(t-t^\prime)$
describes  thermal fluctuations 
with a  strength $T_f$ at   $T_c$.

The typical values of the Ginzburg--Landau parameters for  Fermi liquids
are:
$\tau_{GL} (T_\infty )=\tau _0/(1-T_\infty/T_c )$, 
$ \xi (T_\infty) \sim\xi _0 /(1-T_\infty /T_c)^{1/2}$,
$\xi _0 =\hbar v_F/2\pi T_c $ and $\tau _0=\pi \hbar/8T_c$. 
The diffusion constant 
$\sigma\sim \ell /\xi _0$,  
$\ell $ is the mean free path of a quasiparticle. In $^3$He, 
$\sigma $ is very large because $\ell /\xi _0\sim 10^{3}$.
The noise strength  is 
$
T_f \sim {\rm Gi}^{-1}
\left[1-(T/T_c)\right]^{-1/2} 
$, 
${\rm Gi}=\nu (0)\xi _0^3T_c \sim 10^4$ is the Ginzburg number and
$\nu (0)$ is normal density of states.

{\it Results}. -- 
We  solved Eq. 
(\ref{gle1}) by the 
implicit Crank-Nicholson method. 
The integration domain
was equal to $150^3 $  units of Eq. (\ref{gle1}) with  
$200^3$ mesh points. 
The boundary conditions were taken as $\partial_z \psi = i k \psi$ 
with a constant $k$
at the top and the bottom of the integration domain. This
implies  
a  superflow $j_s = k | \psi|^2$ along the $z$--axis 
far away from the temperature bubble.
The simulations were carried out on a massive parallel 
computer, the  Origin 2000, at Argonne National Laboratory.

\begin{figure}[h]
\centerline{ \psfig{figure=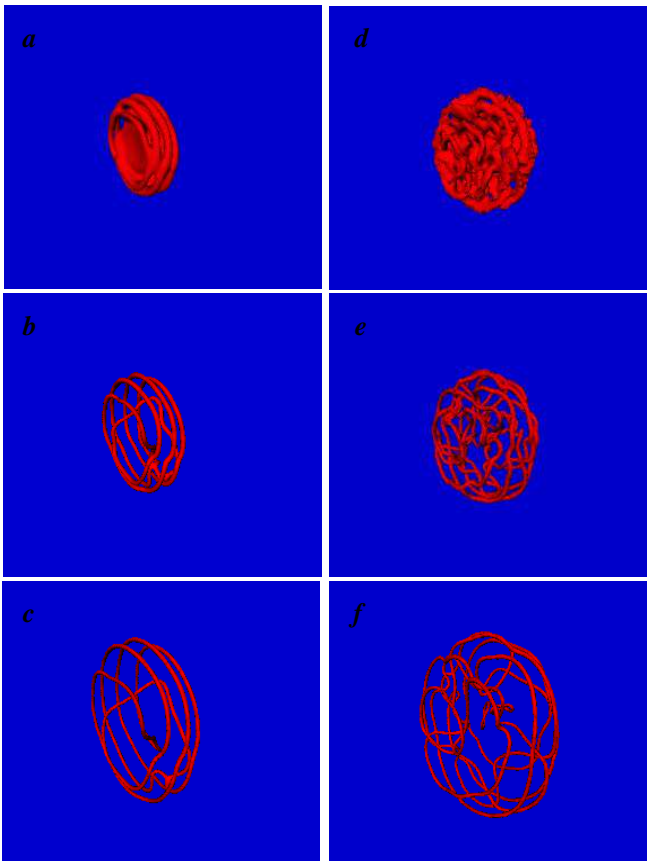,height=4.5in}}
\caption{3D isosurface of
$|\psi|=0.4$ for $\sigma=400, E_0=30$ and $k=0.5$.
(a-c) $T_f=0$. Images are taken at times $t=36,48,80$.
(d-f), $T_f=0.002$, $t=24, 48 , 80$.
}
\label{Fig2}
\end{figure}

Selected  results are shown in Fig. \ref{Fig2}.
One sees (Fig. \ref{Fig2}a-c) that without fluctuations 
(numerical noise only \cite{comment})
 the  vortex rings nucleate upon the passage of the
thermal front. Not all of the rings survive: the 
small ones collapse and only the big ones grow. Although the vortex lines
are centered around the  point of the quench, they exhibit a
certain degree of entanglement. After a long transient period,
most of the vortex rings reconnect and form the 
almost axisymmetric configuration. 

We find that the fluctuations have a  strong effect at
early stages: 
the vortices nucleate not only at the 
normal-superfluid interface, but also in the bulk of the supercooled 
region (Fig. \ref{Fig2}d-e). 
However,  later on, small vortex rings in the interior collapse and only 
larger rings
(primary vortices) survive and expand (Fig. \ref{Fig2}f).

\begin{figure}[h]
\centerline{ \psfig{figure=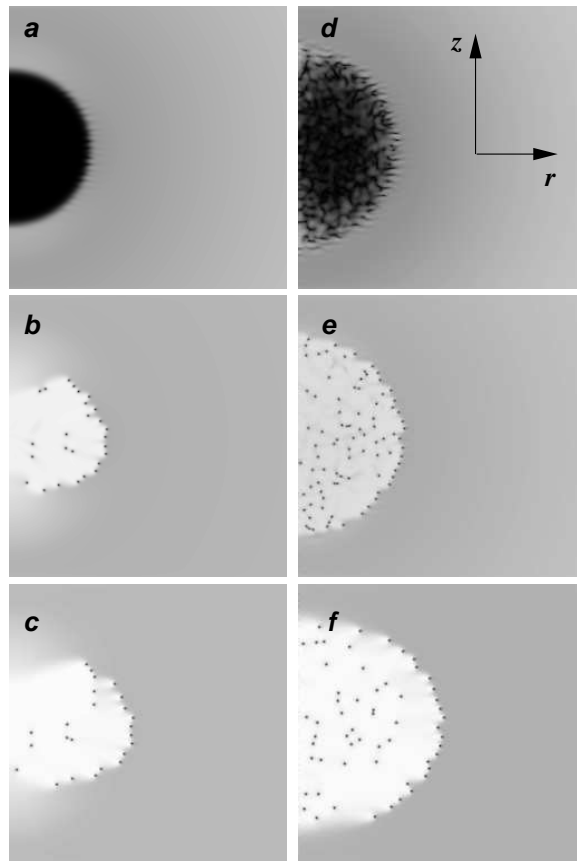,height=4.5in}}
\caption{Grey--coded images of $|\psi|$ for
axi--symmetric  Eq. (\protect\ref{gle1})
for $\sigma=5000$, $E_0=50$ and $k=0.5$, black corresponds to $|\psi|=0$
and white to $|\psi|=1$. Current is along the
$z$-axis. Vortices are seen as black dots.
(a-c) $T_f=0$, images are  shown for $t=40, 100, 200$;
(d-f) $T_f=0.002$,  for $t=30, 50, 200$
}
\label{Fig3}
\end{figure}

\begin{figure}[h]
\centerline{ \psfig{figure=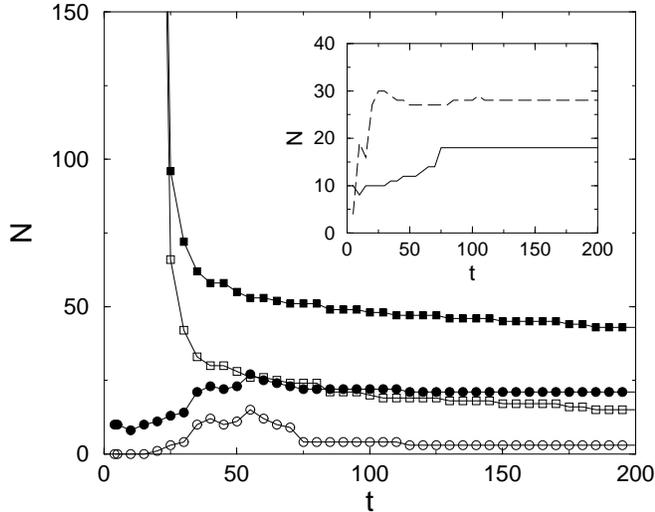,height=2.8in}}
\caption{Number of vortices $N^+$  (filled symbols) and anti-vortices 
$N^-$ (open symbols) vs time for $\sigma=5000, E_0=50$ and $k=0.5$. 
Circles correspond to $T_f=0$ and squares to $T_f=0.002$. Inset: 
$N=N^+-N^-$ for $T_f=0$ (solid line) and $T_f=0.002$ (dashed line)
}
\label{Fig_new}
\end{figure}

To elucidate the details of nucleation we  considered an
axi-symmetric version of Eq. (\ref{gle1})
(depending on only $r$ and $z$ coordinates, $\Delta=
\partial_r^2+1/r \partial_r +\partial_z^2$)
for  realistic $^3$He
parameters \cite{explos}:
$k\sim 1$, $E_0\gg 1$, and $\sigma \sim 10^3$.
The 
domain was $500^2$ with $1000^2$ mesh points. 
We have  found that without thermal 
fluctuations  the vortices
nucleate at the front of the normal-superfluid interface (black/white 
border in  Fig. \ref{Fig3}a-c) analogous to the 3D  case.
The initial instability is seen as a
corrugation of the interface. The interface propagates towards the
center,
leaving the vortices behind. 
As thermal fluctuations are turned on, the vortex rings also nucleate 
in the bulk 
of the supercooled region (Fig. \ref{Fig3}d) resulting in the creation  
of the secondary  vortex/antivortex pairs. 
We have found that  the 
``primary''  vortices prevent the supercurrent 
from penetrating into the region filled with the secondary vortices. 
One sees that the primary 
vortices encircle the brighter spot in 
Fig. \ref{Fig3} indicating a larger 
value of the order parameter and thus a smaller value of the
supercurrent.
 As a result the secondary vortices either annihilate with antivortices
due to their mutual attraction or
 collapse due to the absence of the inflating superflow. 
Fig. \ref{Fig_new} shows the number of
 vortices $N^+$ and antivortices $N^-$ vs 
time with and without fluctuations. Fluctuations initially create 
a very large number $\sim 10 ^4$ of vortices and antivortices
in the bulk  which then annihilate.
The 
resulting amount of surviving vortices $N=N^+ -N^-$ is only weakly
dependent on 
fluctuations.

\begin{figure}[h]
\centerline{ \psfig{figure=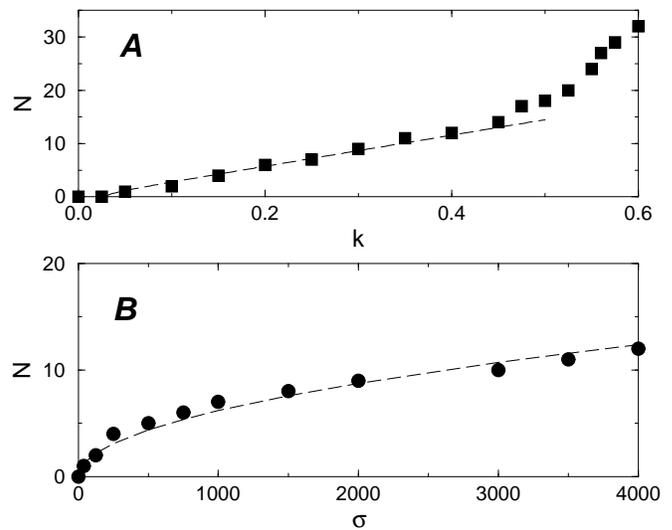,height=2.8in}}
\caption{(a) Number of  vortex rings $N$ as function of $k$
for $E_0=50$ and $\sigma=5000$ and (b) $N$ vs  $\sigma$
for $k=0.4$ and $E_0=50$.
Dashed lines show fitting to prediction Eq. (\protect
\ref{estimate}).
}
\label{Fig4}
\end{figure}

Shown in Fig. \ref{Fig4} is the number of  vortex rings $N$ vs 
quench parameters  and applied current $k$. 
At small $k$ $N$ shows threshold behavior while becoming almost linear 
for larger $k$ values.
The deviations from linear a law appear close to 
the value of the critical current $k_c= 1/\sqrt 3$ for a  homogeneous
system.

{\it Stability of normal-superfluid  interface}.-- 
Following Ref.\cite{zurek,kopnin}, we expand the 
local temperature  $1-f$ near $T_c$. Let us put $x=r_c-r$ 
where $r_c$ is the radius of the surface at which $T=T_c$ or $f=1$, i.e.,
$r_c^2=(3/2)\sigma t\log (t_{\rm max}/t)$. 
A positive $x$ is directed towards the hot region. We write 
$1-f(r ,t) \approx  -\alpha   (x-v t ) $ 
where $\alpha =- [df/dr]_{f=1} = 2 r_c/\sigma t $ is the local
temperature
gradient  
and $v= (\alpha \tau _Q)^{-1}$ is the front velocity defined through 
the quench rate
$\tau _Q^{-1}=\left[\partial f/\partial t\right]_{f=1}$. We have
$
v= \left( 3 \sigma t -2r_c^2\right)/4r_ct 
$. The front starts to move towards the center  at
$t>t_*=t_{\rm max}/e$ and disappears at $t=t_{\rm max}$ when the
temperature
drops below $T_c$. The front velocity accelerates as 
the hot bubble collapses.
Since the front radius $r_c$ is large compared to the coherence length, 
it can be considered flat. We chose the coordinate $y$ parallel to the
front.

We transform to the frame moving with the velocity $v$ and perform
the scaling of variables $ 
\tilde x, \tilde y   = (x, y)v,~   
\tilde t  =  t v^2,~   
\tilde \psi   =  \psi /v$, and $u  =  v^3/ \alpha$. The parameter 
$u\sim (\sigma ^2/t_{\rm max})/\log ^2 (t_{\rm max}/t)$ 
is of the  order 1
in the experiment \cite{explos} at the initial time but grows
rapidly
as the hot bubble shrinks. In our numerical simulations, $u\gg 1$.
Eq. (\ref{gle1}) takes the form          
(we  drop  tildes in what follows)
\begin{equation} 
\partial_t \psi = \Delta \psi +  \partial_x \psi -\frac{x}{u}  
\psi -|\psi| ^2 \psi 
. \label{gle3}
\end{equation}  
The amplitude $F$ of  the 
current-carrying solution  $ \psi = F\exp (iky) $ satisfies 
\begin{equation}
 \partial_x^2  F  +
  \partial_x F -\left(\frac{x}{u} +k^2\right)   F - F^3 = 0.
\label{stat}
\end{equation} 
To examine the transverse 
stability of stationary  solution to Eq. (\ref{stat}) we put
$\psi = (F + w ) \exp(i k x)   $  where the real and imaginary parts of 
the perturbation $w=a+i b$ are 
\begin{equation} 
{a \choose b } = { A \choose i B} \exp (\lambda(q)  t 
+ i q y) 
\label{sel}
\end{equation} 
where $q$ is the transverse undulations wavenumber and $\lambda$ is the 
corresponding growth rate, we obtain
\begin{eqnarray} 
\Lambda A+2 \chi  B  &=& \partial_x^2 A + \partial_x A-\frac{x}{u} 
A-3 F^2 A  \nonumber \\
\Lambda B +2 \chi   A&=& \partial_x^2 B + \partial_x B-\frac{x}{u} B
- F^2 B
\label{lin2} 
\end{eqnarray}
where 
$\chi = k q$ ,  $\Lambda = \lambda + q^2 $,  and $x\to x+u  k^2$. 

The eigenvalue $\Lambda$ for 
$\chi \to 0 $ can be found as an expansion in $\chi$: 
$\Lambda = \chi \Lambda_1 + \chi^2 \Lambda _2 ^2 +...$ and similarly for 
$A$ and $B$. In zeroth order in $\chi$ one has
$A_0=0,  B_0 =  F$.  
In the first order we derive  $B_1=0$ and
\begin{equation} 
\partial_x^2 A_1 + \partial_x A_1-\frac{x}{u}  A_1-3 F^2 A _1=2F .
\label{lin3} 
\end{equation} 
The solution $A_1 = 2 u \partial_x F$ is obtained by 
differentiating Eq. (\ref{stat}).  
In the second order to   Eq.  (\ref{lin2}) one has 
\begin{equation} 
\partial_x^2 B_2  + \partial_x B_2 -\frac{x}{u}  B_2 - F^2 B_2 =
 4u  \partial_x F  +\Lambda_2 F.
\label{lin4} 
\end{equation}
A zero mode of Eq. (\ref{lin4}) is  $F$. The adjoint 
function is $B^+ = F \exp (x)$. Eq. (\ref{lin4}) has a solution if the 
solvability condition with respect to the zero mode  is fulfilled 
\begin{equation} 
\int^{\infty}_{-\infty} dx F e^{x} ( 4 u  \partial_x F  +\Lambda_2 F) =0 
\label{solv} 
\end{equation} 
After integration we obtain  
$\Lambda_2 = 2 u$. 
Returning to the original notations, we obtain the {\it exact} result 
\begin{equation} 
\lambda= q^2( 2 u k^2 -1) + O(q^4) 
\label{lambda1} 
\end{equation}
The instability occurs above the threshold value 
$k_v^2=(2u)^{-1}$ or $k_v^2
\sim \alpha ^{2/3}/u^{1/3}
\sim \sigma ^{-1}\log (t_{\rm max}/t)$ in the Ginzburg--Landau 
units.  Since it is much smaller than the bulk 
critical value $k_c = 1/\sqrt{3}$, it can be exceeded for a very small
superflow.

The eigenvalue $\Lambda$ vs   $\chi$ and $u$ can be derived 
independently in the 
limit  $u\gg 1$ assuming that $\Lambda \sim \chi \sim 1/u\ll 1$.
Substituting $x=\bar x -u \gamma$, where $\gamma $ 
determines the position of the interface, 
we treat the terms containing $\Lambda, \chi $ and $\bar x/u$ as 
perturbations for $\epsilon=1/u \to 0$. 
For $\epsilon=0$ and  $\gamma>1/4$ Eqs. (\ref{stat}) possess a front
solution.
This solution should be matched with its asymptotics for $\bar x>0$, 
and this match
fixes  the value of $\gamma$. As it was shown in Ref. \cite{kopnin}, 
for $u \to \infty$ the matching is possible for $\gamma \to 1/4$.

For $\epsilon=0$ Eqs. (\ref{lin2}) 
have 2  zero modes: $(A,B) =(F_x,0)$ and $(A,B)=(0,F)$. 
The solvability conditions  result in the
characteristic equation for $\Lambda$:
\begin{equation} 
\Lambda^2+\frac{1}{u} c_1 \Lambda - 4 c_2  \chi^2 +\frac{d}{u^2}  =0, 
\label{char}
\end{equation} 
where the coefficients $c_{1,2},d$ are given in the forms of 
integrals of $F$  with the corresponding zero modes
in the interval $-\infty <\bar x < x_0$.  
The constant $x_0$ is determined from the condition 
$d=0$ because 
for $\chi=0$ there is always an exact 
solution to Eq. (\ref{lin2}) with $\Lambda=0$. 
Substitution of the solutions for
 $\gamma \to 1/4 $ yields 
$c_1   \to  2,\; c_2  \to  1 $
and the largest growthrate of transverse perturbations  
\begin{equation}
\lambda = \sqrt{1/u^{2}  + 4 k^2 q^2 } -1/u  -q^2  . 
\label{lambda7}
\end{equation}
Numerical solution of Eqs.  
(\ref{lin2}) demonstrates an excellent
agreement with the theoretical expression  Eq. (\ref{lambda7}).

Now we apply the above results to estimate the 
number of nucleated vortices. 
The evolution of perturbations near the interface  is 
given  by the   integral
\begin{equation} 
w \sim \int dq \exp[\lambda(q) t + iqy ] .
\label{int} 
\end{equation} 
In the case of thermal quench, the normal/superfluid front velocity 
$u\to \infty$ as time increases, therefore, the limit of large $u$
applies. 
For $u\to \infty$ one has  
$\lambda = 2 |k q|  -  q^2 $.
The maximum growth rate is achieved at $q=k$ and is simply $k^2$. 
Taking into account that it is the thermal noise which provides 
initial perturbations for the interface instability, 
one derives from Eqs.(\ref{lambda7},\ref{int})
$\langle |w| \rangle  \sim \sqrt{T_f} \exp [ k^2 t + i ky] $. 
The  number of vortices is estimated as 
$N = r_0 k$, where $r_0$ is the radius of the front where the 
perturbations $w$ become of the  order of one. 
The time interval $t_0$  corresponding to $\langle |w| \rangle =1$ is 
$t_0 \sim k^{-2} \log (T_f^{-1})$. Vortices have no time to grow if
$t_0\to t_{\rm max}$.
For  $r_0$ one then finds:
\begin{equation}
r_0^2 =(3/2)\sigma (t_{\rm max}-t_0) \log  \bigl(t_{\rm max}/
(t_{\rm max}-t_0) \bigr)
. 
\label{tst}
\end{equation}
The number of vortices with logarithmic accuracy is
\begin{equation}
N \sim k r_0  \sim  \sqrt\sigma E_0^{1/3}\sqrt{ \left(v_s/v_c\right)^2 
-\beta^2\log(T_f^{-1})/ E_0^{2/3} } 
\label{estimate} 
\end{equation}
where $\beta=const$, while $v_s$ and $v_c$ are the imposed and critical GL
superflow velocity, respectively. This estimate is in agreement with
the results of simulations, see Fig. \ref{Fig4}. 
Eq. (\ref{estimate}) exhibits a slow logarithmic dependence of the 
number of vortices at the interface 
on the level of fluctuations and agrees with the 
results presented in Fig. 3 \cite{comment1}. 
For  
 $\sigma \sim 10^3, E_0 \sim 10^2-10^3$ which is close to the experimental
values of the parameters. Our analysis results in about 10  
surviving vortices per heating event. It is consistent with Ref. \cite{explos} 
where as many as 6 vortices per neutron were detected. 

{\it In conclusion}, we have found that the rapid normal--superfluid  
transition in the presence of superflow is dominated by a transverse
instability of the normal/superfluid interface propagating from
the bulk into the  normal region. 
This instability produces primary vortex loops 
which then separate from the interface. Simultaneously, a large number of 
vortex/antivortex pairs are created by 
fluctuations in the bulk of the supercooled 
region formed after the collapse of the hot bubble. The primary vortex loops
screen out the superflow and cause annihilation
of the vortex/antivortex pairs in the bulk. The  number of surviving 
vortices is 
determined by  superflow-dependent optimum wavevector 
of the interface instability.

We are grateful to V. Eltsov, M. Krusius,  G. Volovik and W. Zurek  for
stimulating discussions.
This research is supported by
US DOE, grant W-31-109-ENG-38,  and by NSF, STCS \#DMR91-20000.

\end{document}